\documentclass[aip, jcp, reprint,twocolumn,showpacs, amsmath,amssymb,superscriptaddress]{revtex4-1}
\usepackage{graphicx,color}
\usepackage{dcolumn}
\usepackage{bm}
\usepackage{wrapfig}
\usepackage{float}
\usepackage{lipsum}
\usepackage{caption}
\usepackage{natbib}
\usepackage{placeins}
\usepackage{comment}
\usepackage{centernot}
\usepackage{multirow}
\usepackage{mathtools}
\usepackage[utf8]{inputenc}
\usepackage{titlesec}
\usepackage[titletoc,toc,title]{appendix}
\usepackage[caption=false]{subfig}
\usepackage[font=small,labelfont=bf,tableposition=top]{caption}
\usepackage{fancyhdr}
\begin{document}

\title{Exploring and enhancing the accuracy of interior-scaled Perdew-Zunger self-interaction correction }

\author{Puskar Bhattarai}
\email{puskar.bhattarai@temple.edu}
\affiliation{Department of Physics,Temple University, Philadelphia,PA-19122}
\author{Biswajit Santra}
\affiliation{Department of Physics,Temple University, Philadelphia,PA-19122}
\author{ Kamal Wagle}
\affiliation{Department of Physics,Temple University, Philadelphia,PA-19122}
\author{Yoh Yamamoto}
\affiliation{Department of Physics, University of Texas at El Paso, El Paso, Texas 79968, USA}
\author{Rajendra R. Zope}
\affiliation{Department of Physics, University of Texas at El Paso, El Paso, Texas 79968, USA}
\author{Adrienn Ruzsinszky}
\affiliation{Department of Physics,Temple University, Philadelphia,PA-19122}
\author{Koblar A. Jackson}
\affiliation{Department of Physics and Science of Advanced Materials, Central Michigan University, Mount Pleasant, Michigan 48859, USA}
\author{John P. Perdew}
\affiliation{Department of Physics,Temple University, Philadelphia,PA-19122}
\affiliation{Department of Chemistry,Temple University, Philadelphia,PA-19122}
\date{\today}

\begin{abstract}
 The Perdew-Zunger self-interaction correction (PZ-SIC) improves the performance of density functional approximations (DFAs) for the \textcolor{black}{properties that involve} significant \textcolor{black}{ self-interaction error (SIE), as in stretched bond situations, but overcorrects} for equilibrium properties where \textcolor{black}{SIE is insignificant.}
 This overcorrection is often reduced by LSIC, local scaling of the PZ-SIC to the local spin density approximation (LSDA).  
 Here we propose a new scaling factor to use in an LSIC-like approach that satisfies an additional important constraint: the correct coefficient of atomic number $Z$ in the asymptotic expansion of the exchange-correlation (xc) energy for atoms.
 LSIC and LSIC+ are scaled by \textcolor{black}{functions of the iso-orbital} indicator $z_{\sigma}$, which \textcolor{black}{distinguishes} \textcolor{black}{one-electron regions from many-electron regions.} LSIC+ applied \textcolor{black}{to} LSDA works better for many \textcolor{black}{equilibrium} properties than LSDA-LSIC and the Perdew, Burke, and Ernzerhof (PBE) generalized gradient approximation (GGA), and \textcolor{black}{almost as well as} the strongly constrained and appropriately normed (SCAN) meta-GGA.
 LSDA-LSIC and LSDA-LSIC+\textcolor{black}{,}  however, both fail to predict interaction energies involving weaker bonds, in sharp contrast to their earlier successes. It is found that \textcolor{black}{more than one set of localized SIC orbitals can yield a nearly degenerate energetic description of the same multiple covalent bond, suggesting that a consistent chemical interpretation of the localized orbitals requires a new way to choose their Fermi orbital descriptors.} To make a locally scaled-down SIC to functionals beyond LSDA requires a gauge transformation of the functional’s \textcolor{black}{energy density. The resulting SCAN-sdSIC}\textcolor{black}{, evaluated on SCAN-SIC total and localized orbital densities, leads to an acceptable description of many equilibrium properties including the dissociation energies of weak bonds.}

\end{abstract}

\maketitle
\section{Introduction}
 Kohn-Sham density functional theory(KS-DFT)\cite{Kohn} is \textcolor{black}{a computationally efficient} approach to \textcolor{black}{calculate the ground state energy and  density of many-electron systems.} It is widely used \textcolor{black}{to predict} various properties of atoms, molecules and \textcolor{black}{solids,} because of its computational efficiency \textcolor{black}{in comparison with many-electron} wave-function calculations. The accuracy \textcolor{black}{of KS-DFT for ground state calculations} depends upon the approximation \textcolor{black}{for} the exchange-correlation energy ($E_{xc}$) of the system. Jacob's ladder\cite{Jacob} is \textcolor{black}{often used to classify the} approximate density functionals, where higher rungs incorporate additional \textcolor{black}{ingredients that increase both accuracy and computational cost. $E_{xc}$} \textcolor{black}{is often approximated} as 
 \begin{equation}
     E_{xc}[n_\uparrow,n_\downarrow,]=\int d^3r n \epsilon_{xc}(n_\uparrow,n_\downarrow,\nabla n_\uparrow,\nabla n_\downarrow,\tau_\uparrow,\tau_\downarrow),
 \end{equation}
 where
 \begin{equation}
     n(\textbf{r})= n_\uparrow (\textbf{r})+ n_\downarrow(\textbf{r})
 \end{equation}
  is the total density and 
\begin{equation}
    \tau_{\sigma}(\textbf{r})=\sum _i^{occup}\frac{1}{2}|\nabla \psi_{i\sigma}(\textbf{r})|^2
\end{equation}
 is the \textcolor{black}{positive} kinetic energy density for the occupied orbitals $\psi_{i \sigma}$.
 The first three rungs \textcolor{black}{include the local} spin density approximation(LSDA), generalized gradient approximation (GGA) and meta-GGA. LSDA takes the \textcolor{black}{local} electron density as its \textcolor{black}{sole} ingredient. GGAs \textcolor{black}{take the gradient} of electron density $\nabla n_\sigma$ \textcolor{black}{as an additional ingredient, and meta-GGAs} take $\tau_{\sigma}$ \textcolor{black}{as well}. \textcolor{black}{Semi-local} non-empirical approximate functionals of higher rungs are often constructed by satisfying more exact constraints and norms than the lower rungs.\cite{SCAN} For instance, \textcolor{black}{the} PBE\cite{PBE} GGA satisfies 11 known exact constraints and \textcolor{black}{the} SCAN\cite{SCAN} meta-GGA satisfies 17 known exact constraints. The most common norm is the uniform electron gas,\cite{PW92} for which almost all the functionals are exact by construction.
    
  All semi-local \textcolor{black}{DFAs} suffer from self-interaction error (SIE)\cite{PZ81} due to the imperfect cancellation of \textcolor{black}{the self-Hartree energy $U$} by \textcolor{black}{the self-exchange-correlation energy $E_{xc}$} of a single fully occupied orbital. \textcolor{black}{The} \textcolor{black}{Perdew-Zunger} self-interaction correction (PZ-SIC)\cite{PZ81} removes one-electron SIE by subtracting it on an \textcolor{black}{orbital-by-orbital} basis. Hence, the corrected $E_{xc}$ can be written as 
  \vspace{0.15cm}
  \begin{equation}
      E_{xc}^{PZ-SIC}=E_{xc}^{approx}[n_\uparrow,n_\downarrow]-\sum_{\alpha\sigma}\delta_{\alpha \sigma}.
      \vspace{0.15cm}
  \end{equation}
  \textcolor{black}{Here} $\delta_{\alpha \sigma}$ is the SIE of \textcolor{black}{an} orbital with quantum numbers $\alpha$ and $\sigma$, (\textcolor{black}{where} $\sigma$ is the spin, and $\alpha$ is the orbital quantum number)\textcolor{black}{. The orbital density is $n_{\alpha\sigma}(\textbf{r})=|\psi_{\alpha\sigma}(\textbf{r})|^2,$ and}
\begin{equation}
 \delta_{\alpha \sigma}=U[n_{\alpha\sigma}]+E_{xc}^{approx}[n_{\alpha\sigma},0].  
  \vspace{0.15cm}
\end{equation} 

Use of delocalized KS orbitals for PZ-SIC \textcolor{black}{leads to a size-extensivity} problem.\cite{PZ81} Besides \textcolor{black}{that,} they can be highly \textcolor{black}{noded, leading to disastrous results}\cite{Stretched} when applied to the semi-local functionals. \textcolor{black}{Fermi-L\"{o}wdin} orbitals (FLOs)\cite{FLOSIC} correspond \textcolor{black}{to a} unitary transformation of KS orbitals and are guaranteed to be localized \textcolor{black}{around} Fermi-orbital descriptor (FOD) positions. FLOs are not nodeless but \textcolor{black}{more weakly} noded than the KS orbitals, as the overlapping real FLOs must have nodes to establish orthogonality. \textcolor{black}{Complex FLOs \textcolor{black}{(not yet implemented)} can achieve orthogonality without \textcolor{black}{having noded orbital densities,} but \textcolor{black}{the orbital densities} may still be lobed and pose problems for semi-local functionals.\cite{Hofmann,Puskar}}

\textcolor{black}{The FLOSIC} method\cite{FLOSIC,self_consistency} has been applied to different density functional approximations (DFAs) to successfully predict various properties of \textcolor{black}{of atoms and molecules}\cite{FLOSIC_app1,FLOSIC_app2,FLOSIC_app3,FLOSIC_app4,FLOSIC_app5,FLOSIC_app6,FLOSIC_app7}. Applying FLOSIC to the approximate semi-local functional greatly reduces the error for stretched bonds\cite{Stretched} where SIE is \textcolor{black}{large,} but it introduces other errors due to the presence of nodes and lobed structures in the orbital densities, which \textcolor{black}{are not needed by} uncorrected DFAs. Besides that, the application of PZ-SIC to the DFAs violates the most common norm of exactness \textcolor{black}{for the} uniform electron gas\cite{How_wrong}\textcolor{black}{,} and produces a significant error \textcolor{black}{in} the exchange-correlation energy of neutral atoms \textcolor{black}{in} the limit of large atomic number $Z$. 
\textcolor{black}{PZ-SIC, although it} improves the barrier heights of chemical reaction\textcolor{black}{s}\cite{Stretched} that \textcolor{black}{include} stretched bonds, often \textcolor{black}{worsens equlibrium properties} like atomization energies\cite{Stretched}, electron affinities, ionization potential, and bond lengths of molecules\cite{Puskar}. 

Various scaling functions and methods\textcolor{black}{\cite{Vydrov,Vydrov_scuseria,Many_electron,Many_electron_1,Klupfel,Puskar,Zope,SOSIC}} have been introduced to \textcolor{black}{improve the prediction of} equilibrium properties like atomization energy. 
The half-SIC approximation proposed in Ref. \onlinecite{Klupfel} \textcolor{black}{does not} perform better for both atomization energy and the energy barriers. The scaling proposed in Ref. \onlinecite{Vydrov} and Ref. \onlinecite{Puskar} violates the correct asymptotic behavior $\frac{-1}{r}$ of the exchange-correlation potential and produces $\frac{-X_{HO}}{r}$ behavior, where $X_{HO}$ is the scaling factor for the highest occupied orbital. Recently, a new approach called LSIC\cite{Zope} was proposed as an interior scaling of PZ-SIC. \textcolor{black}{It} uses \textcolor{black}{an} iso-orbital indicator$(z_\sigma$) as the scaling factor \textcolor{black}{to distinguish one-electron regions from many electron regions.} $z_\sigma=0$ \textcolor{black}{is consistent with a region of uniform density} and  $z_\sigma=1$ \textcolor{black}{is consistent with a one-electron region. $z_\sigma$ is defined by}
 \begin{equation}
     z_\sigma=\frac{\tau_\sigma^W(\textbf{r})}{\tau_\sigma(\textbf{r}),}
 \end{equation}
 where, 
 $\tau_\sigma^W(\textbf{r})$ is the von Weizs\"{a}cker kinetic energy density
 \begin{equation}
 \begin{split}
     \tau_\sigma^W(\textbf{r})=\frac{|\nabla n_\sigma|^2}{8n_\sigma}. \\
     \end{split}
     \label{eq:tau_sigma}
 \end{equation}
 LSIC uses the simplest scaling factor $z_\sigma$ to scale down PZ-SIC and \textcolor{black}{is exact for the} uniform gas limit as satisfied by the uncorrected DFAs. \textcolor{black}{ LSDA-LSIC, evaluated non-self-consistently on LSDA-SIC FLO densities, improves many calculated properties over PZ-SIC, including covalent binding energies\cite{L}.} However, it fails to recover \textcolor{black}{all of the known asymptotic expansion of $E_{xc}$ for atoms of large atomic number $Z$.}
 
 The exact large-$Z$ asymptotic expansions of $E_{xc}$\citep{Pittalis,Julian,Eliott,Cancio} is given as,
 \vspace{0.15cm}
\begin{equation}
    E_{xc}=-A_xZ^{\frac{5}{3}}-A_cZln(Z)+B_{xc}Z+C_xZ^{\frac{2}{3}}+....,
    \label{eq:B_xc}
\vspace{0.15cm}
\end{equation}
where LSDA, PBE and almost all other DFAs \textcolor{black}{exactly reproduce} $A_x$ and $A_c.$\cite{How_wrong} LSDA-sdSIC and PBE-sdSIC\cite{Puskar} do not recover the coefficient $B_{xc}$\textcolor{black}{, but, as will be shown below, SCAN-sdSIC} recovers the leading coefficients $A_x$,  $A_c$, and $B_{xc}$ and produces a balanced result. The relevance of the large-$Z$ expansion to valence-electron properties is discussed in Ref. \onlinecite{Kaplan}. SCAN-sdSIC improves the perfomance of SCAN where SIEs are important and restores \textcolor{black}{most of} the accuracy of SCAN for equilibrium properties, which are severely degraded by SCAN-SIC.\cite{Puskar} SCAN-sdSIC violates the correct asymptotic behavior $\frac{-1}{r}$ of the exchange-correlation \textcolor{black}{potential}. 

We present here the scaling function $f(z_\sigma)$ applied to PZ-SIC\textcolor{black}{, in an approximation referred to as LSIC+,} that recovers the leading coefficients $A_x$,  $A_c$, and $B_{xc}$ and also \textcolor{black}{retains} the correct asymptotic behavior of the exchange-correlation potential. LSIC+ \textcolor{black}{is exact for the} uniform gas limit and produces much less error \textcolor{black}{in} $E_{xc}$ of the neutral atoms \textcolor{black}{in} the limit of large $Z$. LSIC+ satisfies at least one more constraint and is expected to work \textcolor{black}{well} for both equilibrium \textcolor{black}{and stretched-bond properties.} 


\section{Theory and computational details}
\begin{figure}
    \centering
    \includegraphics[scale=0.25]{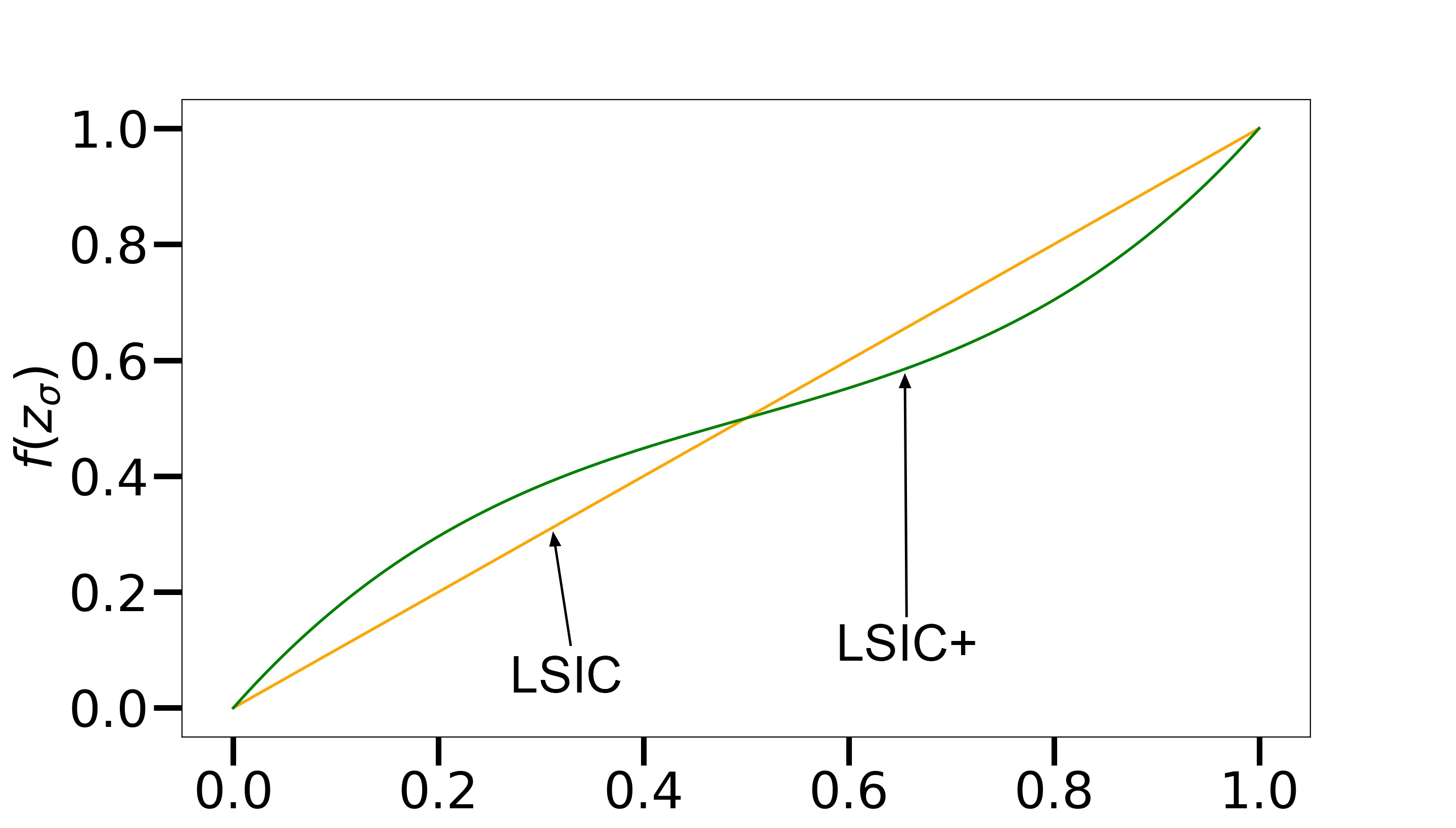}
    \caption{Scaling functions \textcolor{black}{for} LSIC, and LSIC+ as \textcolor{black}{functions} of $z_\sigma$.}
    \label{fig:scaling_function}
\end{figure}

The scaled down SIC using LSIC+ is given as  
\begin{equation}
\begin{split}
    E_{xc}^{DFA-LSIC+}=E_{xc}^{DFA}-\sum^{occ}_{\alpha,\sigma}\{ U^{LSIC+}[n_{\alpha\sigma},0]+\\ E_{xc}^{LSIC+}[n_{\alpha\sigma},0]  \},
    \end{split}
    \label{eq:LSIC+}
\end{equation}
where \begin{equation}
\begin{split}
    U^{LSIC+}[n_{\alpha\sigma},0]=\frac{1}{2}\int\textbf{dr}f(z_\sigma\textcolor{black}{(\textbf{r})})n_{\alpha,\sigma}(\textbf{r})\\ \int\textbf{dr$'$}\frac{n_{\alpha,\sigma}(\textbf{r$'$})}{|\textbf{r}-\textbf{r$'$}|,}
    \end{split}
\end{equation}
\begin{equation}
\begin{split}
    E_{xc}^{LSIC+}[n_{\alpha\sigma},0]=\frac{1}{2}\int  \textbf{dr}f(z_\sigma\textcolor{black}{(\textbf{r})})n_{\alpha,\sigma}(\textbf{r}) \\ \epsilon_{xc}^{DFA}([n_{\alpha,\sigma},0],\textbf{r}).
    \end{split}
\end{equation}
$\epsilon_{xc}^{DFA}$ is the exchange-correlation energy density per electron\textcolor{black}{,} and $f({z_\sigma})$ is the scaling factor such that $f(0)=0$, $f(1)=1$, i.e.\textcolor{black}{,} PZ-SIC is scaled in such a way that there is full correction for \textcolor{black}{any one-electron} orbital density and no correction for \textcolor{black}{any} uniform electron density. For LSIC+, we propose $f(z_\sigma)$ as 
\begin{equation}
    f(z_\sigma)=\frac{1}{2}+a(z_\sigma-\frac{1}{2})+b(z_\sigma-\frac{1}{2})^3
\end{equation}
where, $b=4(1-a)$. The \textcolor{black}{value of $a$ is chosen \textcolor{black}{to recover} the correct coefficient $B_{xc}$ of the large-$Z$} asymptotic expansion of \textcolor{black}{Eq. \ref{eq:B_xc}}. $a=0.5$, and $b=2.0$ make the large-$Z$ asymptotic expansion very close to the exact one. \textcolor{black}{Table \ref{tab:Bxc} shows the $B_{xc}$ values for LSIC, LSIC+ and SCAN-sdSIC. \textcolor{black}{They are based on calculations} performed for four rare gas atoms Neon($Z=10$), Argon ($Z=18$), Krypton ($Z=36$), and Xenon ($Z=54$), and extrapolated to $Z\rightarrow \infty$. Here we chose the closed shell atoms to extrapolate and predict the coefficients \textcolor{black}{in order to minimize} the effect of shell structure as discussed in Ref. \onlinecite{Cancio}. Fig. \ref{fig:scaling_function} shows the scaling functions of LSIC and LSIC+ plotted as a function of $z_\sigma$. LSIC+ produces more correction than LSIC in the region 0f $0.0 < z_\sigma \leq 0.5$ and less correction than LSIC in the region $0.5 \leq z_\sigma < 1.0 $.}  

\textcolor{black}{All calculations, including \textcolor{black}{$E_{xc}$ of rare gas atoms, \textcolor{black}{were}} carried out} using a \textcolor{black}{developmental} version of \textcolor{black}{the} FLOSIC code\cite{FLOSIC-code} based on \textcolor{black}{the} NRLMOL code.\cite{NRLMOL} We performed spin unpolarized calculations for rare gas atoms and the S22 set.\cite{S22} All the other calculations are spin polarized. We used the Sadlej basis set\cite{Sadlej} that includes long-range functions to capture the extended nature of anion orbitals for calculating electron affinity. Optimized all-electron Gaussian basis sets\cite{NRLMOL} \textcolor{black}{were} used for all other properties. The FODs \textcolor{black}{were} optimized for SIC calculations until the maximum component of force is less than $5 \times 10^{-4}$ Hartree/Bohr with self-consistent field (SCF) convergence set to $10^{-6}$ Hartree. \textcolor{black}{Unless otherwise stated, all DFA and DFA-SIC calculations (where DFA = LSDA, PBE, or SCAN) are self-consistent, while all DFA-sdSIC, LSDA-LSIC, and LSDA-LSIC+ calculations are performed as a single,non-SCF step using the optimized DFA-SIC density and Fermi orbital descriptors (FODs).}

\textcolor{black}{Full self-consistency requires the \textcolor{black}{complicated implementation of} the functional derivative of $f(z_\sigma)$.} Therefore, we used a quasi-SCF version of LSDA-LSIC and LSDA-LSIC+ to obtain the \textcolor{black}{eigenvalue} of the highest occupied molecular orbital (HOMO) for calculating the vertical electron affinity. A more detailed description of quasi-self-consistent LSDA-LSIC and LSDA-LSIC+ is \textcolor{black}{presented} in \textcolor{black}{Refs.} \textcolor{black}{\onlinecite{quasi,Santosh,FLOSIC_app7}}, which \textcolor{black}{argue} that the main effect of scaling comes from the scaled potential term rather than the variation of the scaling factor, \textcolor{black}{which would} have a minor effect on the results. The quasi-self-consistent version approximates and scales \textcolor{black}{the} PZ-SIC potential as 
\begin{equation}
\label{eq:quasi}
v_{i\sigma}^{interior \ scaled \ SIC} \approx -f(z_\sigma(\textbf{r}))\frac{\delta \{U[n_{i\sigma}]+E_{xc}[n_{i\sigma},0]\}}{\delta n_{i\sigma}(\textbf{r})}.
\end{equation}
\begin{table}[htbp]
  \centering
  \caption{Coefficient of the third term of the large-$Z$ asymptotic expansion \textcolor{black}{of Eq. \ref{eq:B_xc}} using \textcolor{black}{the} LSDA-LSIC+, LSDA-LSIC, and SCAN-sdSIC \textcolor{black}{methods.}}
    \begin{tabular}{ccccc}
    \hline
    \hline
          & $LSIC+$ & LSIC & SCAN-sdSIC  & Exact \\
    \hline
    $B_{xc}$ & -0.1806 & -0.0828 &  -0.1755   & -0.1868\cite{Cancio} \\
    \hline
    \hline
    \end{tabular}
  \label{tab:Bxc}
\end{table}
\begin{figure}
    \centering
    \includegraphics[scale=0.23]{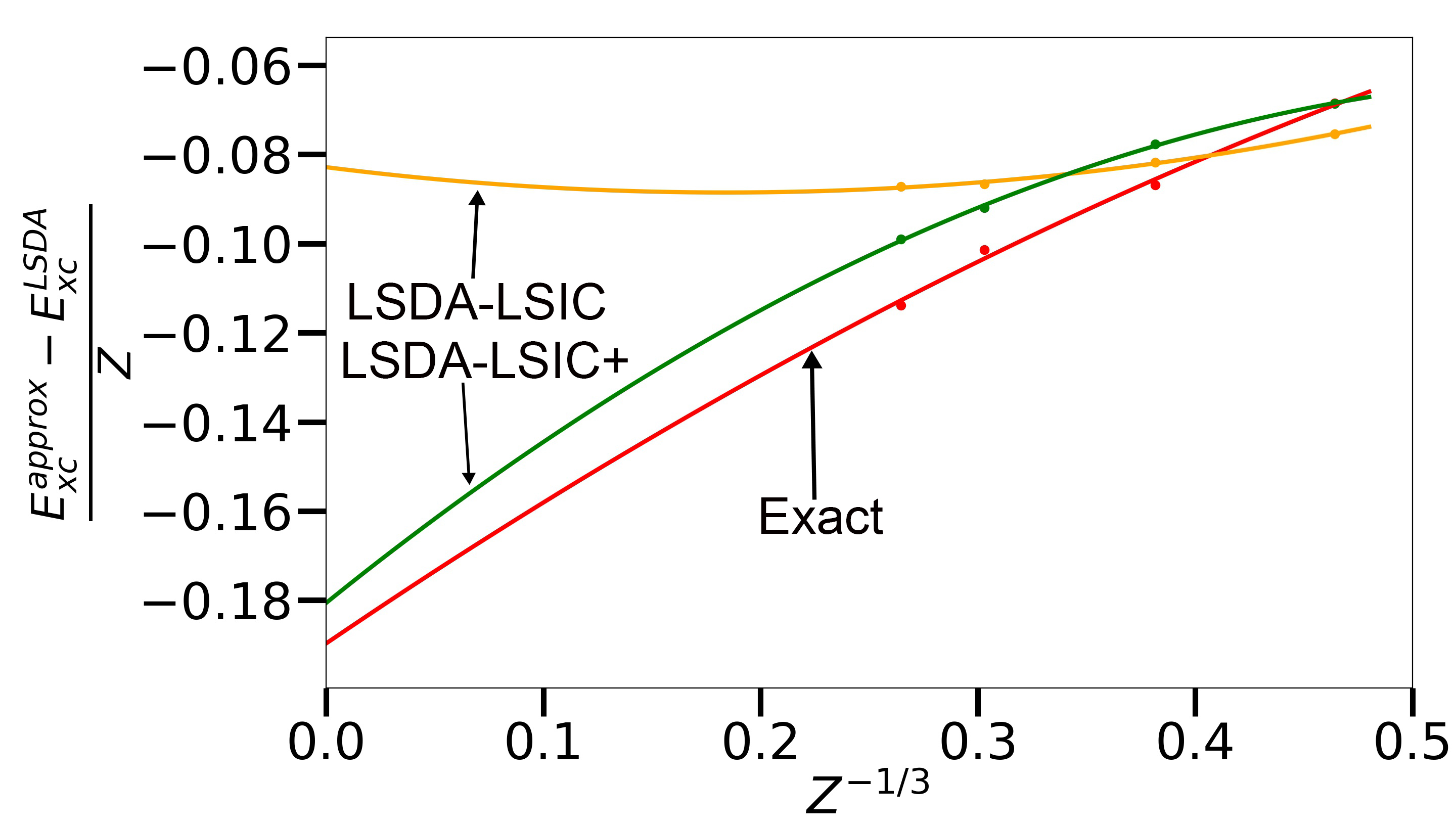}
    \caption{\textcolor{black}{Extrapolation to $Z^{-1/3}\rightarrow 0$ for the coefficient $B_{xc}$ of the large-$Z$} asymptotic expansion of \textcolor{black}{Eq. \ref{eq:B_xc} for} rare gas atoms Neon($Z=10$), Argon ($Z=18$), Krypton ($Z=36$), and Xenon ($Z=54$), (in Hartree) against $Z^{-1/3}$. The solid lines represent the extrapolated curves and the solid dots represent the calculated values at $Z=10, 18, 36,$ and $54.$ Note that $A_x$ and $A_c$ are exact in LSDA.}
    \label{fig:B_xc}
\end{figure}

The LSDA $E_{xc}$ \textcolor{black}{values} of rare gas atoms calculated with LSDA-SIC density are subtracted from exact \cite{Becke,Pittalis}, LSDA-LSIC, and LSDA-LSIC+  \textcolor{black}{$E_{xc}$ values}, and \textcolor{black}{the result is divided by the} atomic number $Z$. \textcolor{black}{The results thus obtained are} plotted against $Z^{{-1}/{3}}$ \textcolor{black}{as} shown in Fig. \ref{fig:B_xc}.  The intercept on the y-axis provides the $B_{xc}$ results. The oscillation of the LSDA exchange curve \textcolor{black}{imitates that of the exact} exchange energy curve. \textcolor{black}{Subtracting the LSDA values minimizes} the oscillations due to the shell structure.\cite{Eliott} Fig. \ref{fig:B_xc} shows that the curve for LSDA-LSIC+ lies very close to the exact curve whereas LSDA-LSIC deviates too much from the exact curve, especially in the higher Z region. We can observe that $B_{xc}$ for LSIC+ is very close to the exact one, \textcolor{black}{as} presented in Table \ref{tab:Bxc}.

\begin{figure}[H]
    \centering
    \includegraphics[scale=0.23]{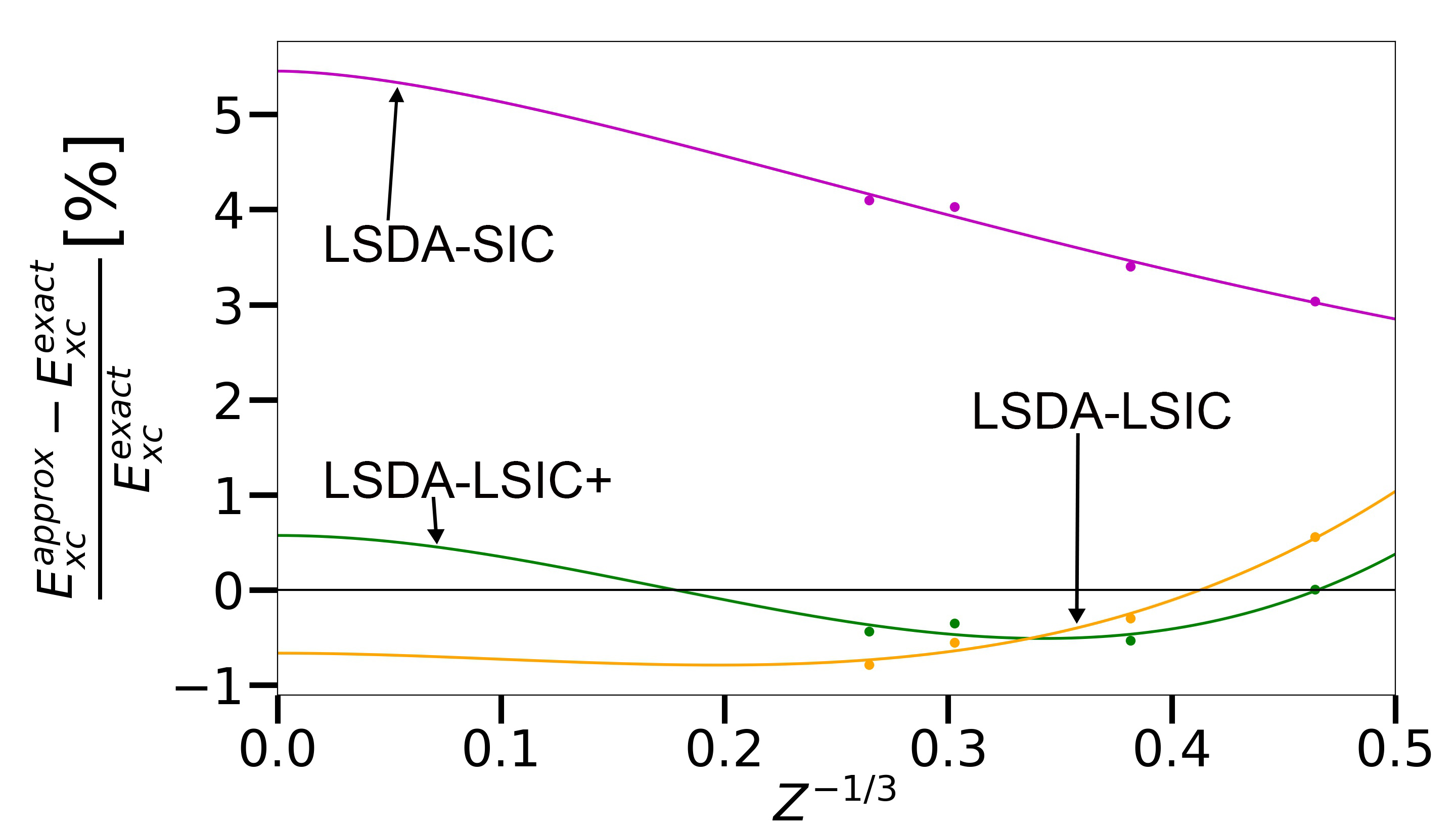}
    \caption{Extrapolation \textcolor{black}{as in Ref. \onlinecite{How_wrong}} of percentage error of $E_{xc}$ \textcolor{black}{for} rare gas atoms against $Z^{-1/3}$. The solid lines represent the extrapolated curves and the solid dots represent the calculated values at $Z=10, 18, 36,$ and 54.}
    \label{fig:error_percent}
\end{figure}

Fig. \ref{fig:error_percent} shows the relative percentage error of $E_{xc}$ with LSDA-SIC, LSDA-LSIC, and LSDA-LSIC+ \textcolor{black}{and it demonstrates that LSDA-LSIC and LSDA-LSIC+} \textcolor{black}{respect} the norm of the uniform electron gas\textcolor{black}{, since the} percentage error for \textcolor{black}{exchange-correlation energy becomes nearly zero} in the limit of large $Z$. \textcolor{black}{These extrapolated errors are} slightly \textcolor{black}{different from zero since} the extrapolation is performed taking just four points \textcolor{black}{into} account. For $Z\rightarrow \infty$, LSDA-SIC has the highest error of 5.46\%, LSDA-LSIC has an error of -0.67\% and LSDA-LSIC+ has an error of 0.57\%. The error of these methods for $E_{xc}$ of rare-gas atoms is shown in \textcolor{black}{table} \ref{tab:Exc_error}.
\\
\begin{table}[htbp]
  \centering
  \caption{Mean percentage error(MPE) and mean absolute percentage error(MAPE) of $E_{xc}$ \textcolor{black}{for} \textcolor{black}{four rare gas atoms Neon($Z=10$), Argon ($Z=18$), Krypton ($Z=36$), and Xenon ($Z=54$).} }
    \begin{tabular}{cccc}
    \hline
    \hline
          & LSIC+ & LSIC  & LSDA-SIC \\
          \hline
    MPE   & -0.3301 & -0.2721 & 3.6418 \\
    MAPE  & 0.3324 & 0.5492 & 3.6418 \\
    \hline
    \hline
    \end{tabular}%
  \label{tab:Exc_error}%
\end{table}%

\section{\textcolor{black}{Results and Discussion}}
LSIC+ scaling is applied to determine different ground-state properties. We compare several properties of atoms and molecules calculated with LSDA, PBE, and SCAN, with and without SIC, scaled-down SIC(sdSIC)\cite{Puskar}, and the interior-scaling methods LSIC and LSIC+ with LSDA.  The interior-scaling methods with PBE and \textcolor{black}{especially} SCAN do not produce better \textcolor{black}{results} due to gauge inconsistency\textcolor{black}{,} which is discussed in detail in Ref. \onlinecite{Puskar}. The results for PBE-LSIC and SCAN-LSIC are also presented here for comparison purposes only. We will discuss the results for the total \textcolor{black}{energies} of atoms, atomization energies of molecules, barrier heights of \textcolor{black}{chemical reactions}, \textcolor{black}{ionization potentials and electron affinities of atoms and molecules, equilibrium bond lengths of molecules, and the interaction energies} of a few organic complexes. LSIC and LSIC+ are applied \textcolor{black}{to LSDA, but they produce results close to SCAN results and sometimes even better.} Because of the predictive power and success of SCAN\cite{SCAN_success1,SCAN_success2,SCAN_success3,SCAN_success4, SCAN_success5,SCAN_success6,FLOSIC_app6}, the LSDA-LSIC and LSDA-LSIC+ results are primarily compared with SCAN, SCAN-SIC\textcolor{black}{\cite{SCAN-SIC}}, and SCAN-sdSIC results.
\subsection{Atoms}
\begin{figure}[H]
    \centering
    \includegraphics[scale=0.23]{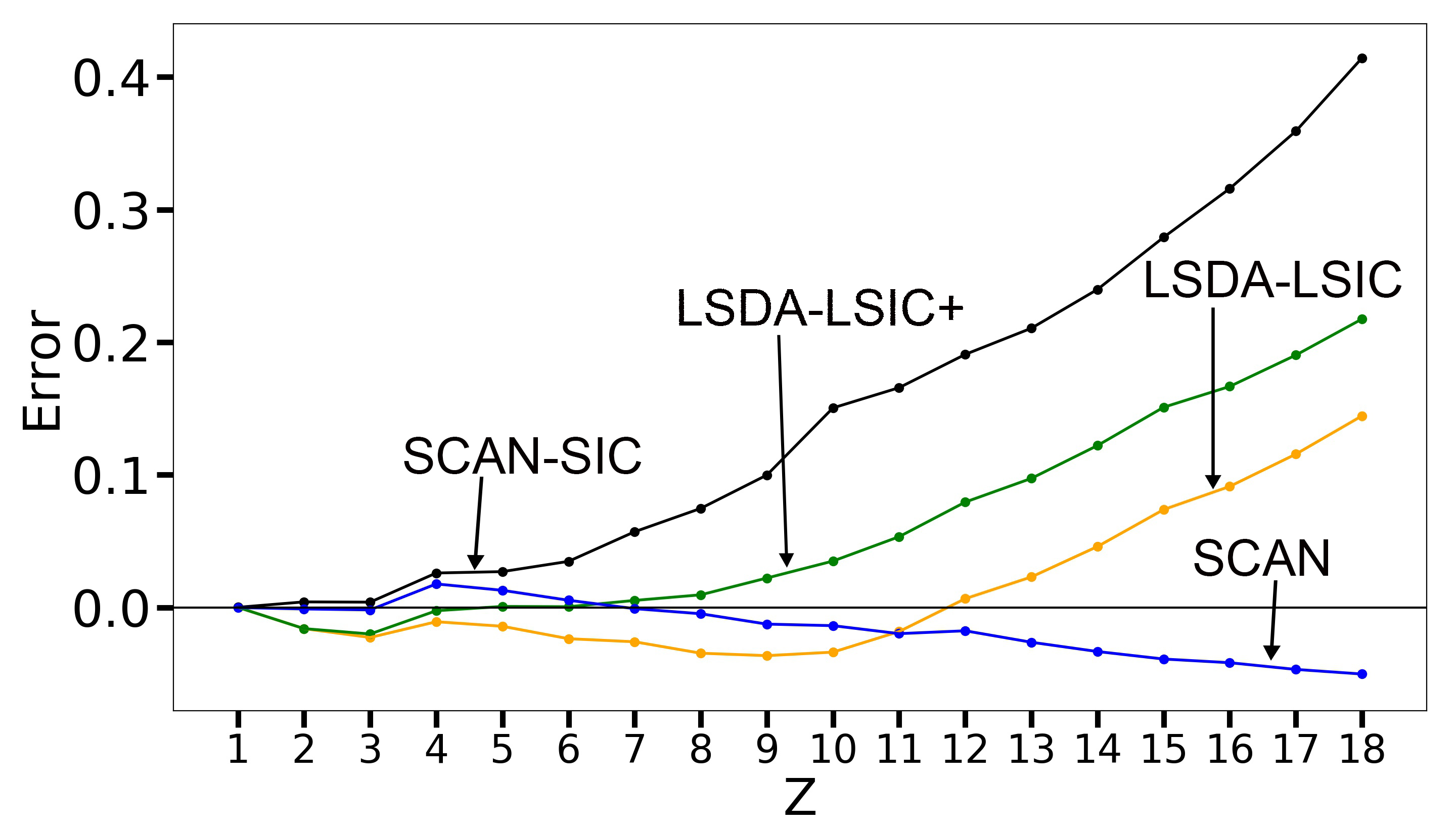}
    \caption{Error (in Hartree) in the total ground state energy plotted against atomic number $Z$ for atoms from H to Ar.}
    \label{fig:atoms_error}
\end{figure}

The mean error (ME), mean absolute error (MAE) \textcolor{black}{,}  and mean absolute percentage error (MAPE) for the total ground state energy of atoms ($Z=1-18$) using different DFAs, their SIC counterparts\textcolor{black}{,} and the scaling methods are presented in Table \ref{tab:Atoms}.
\textcolor{black}{Both LSDA-LSIC and LSDA-LSIC+ significantly reduce the error produced by LSDA-SIC. However, LSIC+ generates slightly more error than LSIC. SCAN serves as the best functional in estimating the total ground state energy of atoms. Fig. \ref{fig:atoms_error} shows the performances of LSDA-LSIC, LSDA-LSIC+, SCAN, and SCAN-SIC in predicting the total ground state energy of atoms. Both LSDA-LSIC and LSDA-LSIC+ produce an improved result for $Z=$1-10 and generates the results closer to the reference values\cite{atoms_ref}. The error for both LSDA-LSIC and LSDA-LSIC+ rises rapidly as we move from $Z=$11 to $Z$=18.}

\begin{table}[htbp]
  \centering
  \caption{ ME (in Hartree), MAE (in Hartree), and MAPEs for total ground state \textcolor{black}{energies} of atoms for atomic number $Z=1$ to $Z=18$ for various levels of approximation.}

\textcolor{black}{
    \begin{tabular}{cccc}
    \hline
    \hline
    Method & MAE   & ME    & MAPE \\
    \hline
    LSDA  & 0.7261 & 0.7261 & -1.0014 \\
    LSDA-SIC & 0.3808 & -0.3808 & -0.2625 \\
    LSDA-sdSIC & 0.0423 & -0.0412 & -0.1081 \\
    LSDA-LSIC & 0.0409 & 0.0147 & -0.0753 \\
    LSDA-LSIC+ & 0.0661 & 0.0618 & -0.0694 \\
          &       &       &  \\
    PBE   & 0.0830 & 0.0830 & -0.0995 \\
    PBE-SIC & 0.1585 & 0.1585 & -0.1091 \\
    PBE-sdSIC & 0.0710 & 0.0710 & -0.0679 \\
          &       &       &  \\
    SCAN  & 0.0192 & -0.0152 & -0.0217 \\
    SCAN-SIC & 0.1474 & 0.1474 & -0.0943 \\
    SCAN-sdSIC & 0.0334 & 0.0334 & -0.0446 \\
    \hline
    \hline
    \end{tabular}%
    }
  
  \label{tab:Atoms}%
\end{table}%

\subsection{Atomization energy}

\textcolor{black}{The} atomization energy of a  molecule is the energy required to break it into its constituent atoms. Table \ref{tab:AE6} presents the MEs, MAEs and MAPEs of \textcolor{black}{the} AE6 set\cite{Lynch} that comprises atomization \textcolor{black}{energies} of six representative molecules. Since the reference atomization energies range from 101 kcal/mol to 1149 kcal/mol, it is reasonable to compare percentage error rather than absolute error.  

\begin{equation}
\label{eq:percentage_error}
\begin{split}
    {\rm Percentage \ error=\frac{calculated \  value-reference \ value}{reference \ value}}\\  \times \ 100\%
    \end{split}
\end{equation}
\begin{table}[htbp]
  \centering
  \caption{ME \textcolor{black}{in kcal/mol}, MAE \textcolor{black}{in kcal/mol} and MAPE for atomization energy of the AE6 set for various levels of approximation.}
    \begin{tabular}{p{10em}ccc}
    \hline
    \hline
    {Approx.} & ME    & MAE   & MAPE \\
    \hline
    LSDA  & 75.5  & 75.5  & 16.7 \\
 
    LSDA-SIC & 53.5  & 57.8  & 9.9 \\
     
    LSDA-sdSIC & 24.5  & 25.7  & 5.6 \\
     
    LSDA-LSIC & -0.9  & 9.3   & 3.0 \\
     
    LSDA-LSIC+ & -0.4  & 8.2   & 2.3 \\
     
    \multicolumn{1}{c}{} &       &       &  \\

    PBE   & 10.6  & 13.8  & 3.8 \\
     
    PBE-SIC & -15.6 & 17.8  & 5.5 \\
     
    PBE-sdSIC & 6.8   & 11.7  & 4.1 \\
     
    \multicolumn{1}{c}{} &       &       &  \\
     
    SCAN  & 0.3   & 3.0   & 1.4 \\
     
    SCAN-SIC & -24.4 & 26.1  & 7.0 \\
     
    SCAN-sdSIC & -3.3  & 5.7   & 2.4 \\
    \hline\hline
    \end{tabular}%
  \label{tab:AE6}%
\end{table}%

\begin{figure}[htbp]
    \centering
    \includegraphics[scale=0.23]{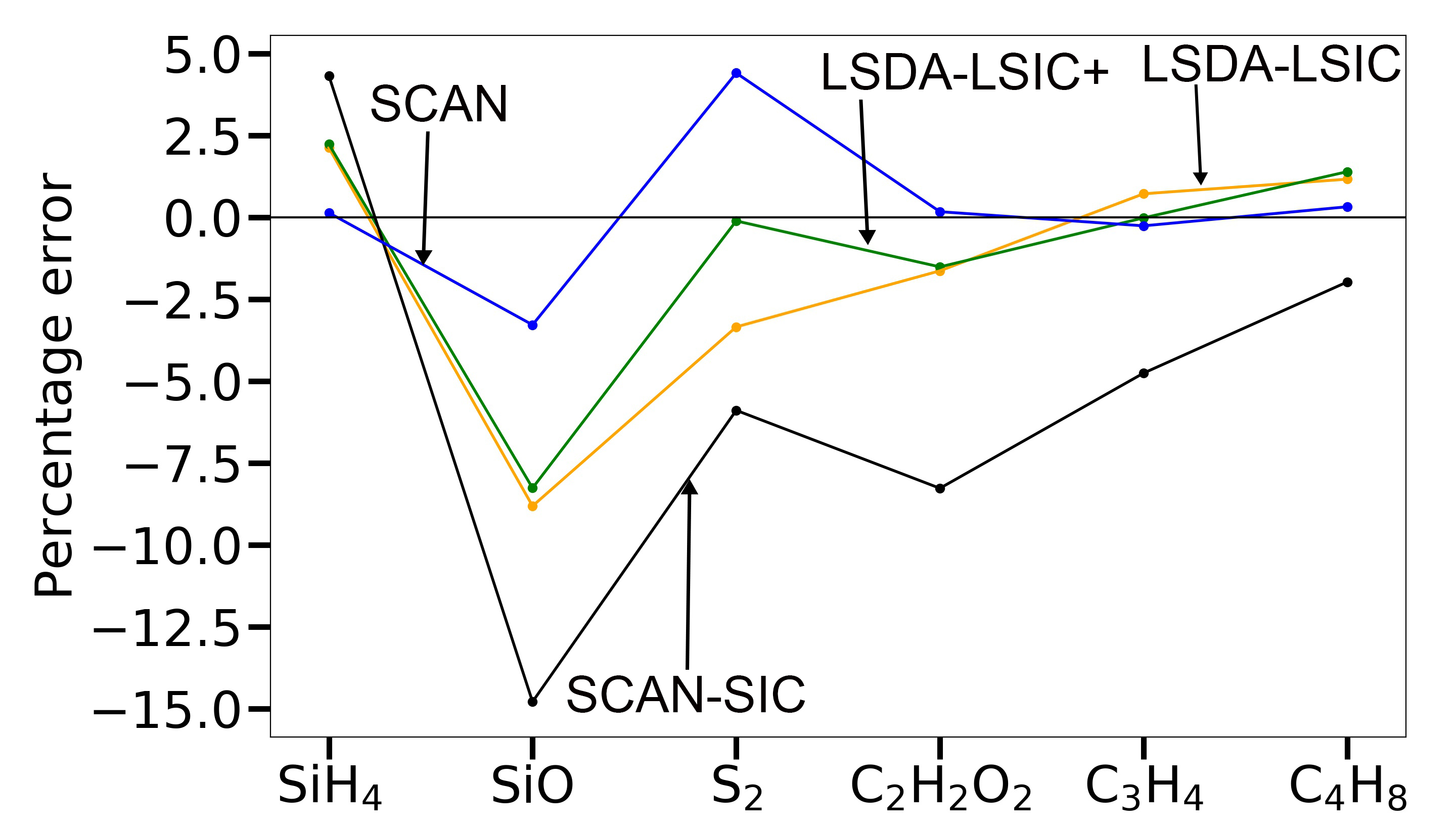}
    \caption{Percentage errors \textcolor{black}{in} \textcolor{black}{atomization energies for} individual molecules of \textcolor{black}{the} AE6 set by SCAN, SCAN-SIC, LSDA-LSIC, and LSDA-LSIC+. }
    \label{fig:AE6_individual}
\end{figure}

LSDA-LSIC+ works better than the LSDA, PBE, their SIC counterparts\textcolor{black}{,} and even LSDA-LSIC. The MAPE of LSDA-LSIC+ is close to SCAN and SCAN-sdSIC.\cite{Puskar} Fig. \ref{fig:AE6_individual} shows the comparison of percentage errors for SCAN, SCAN-SIC, LSDA-LSIC\textcolor{black}{,}  and LSDA-LSIC+ for the individual molecules. Percentage errors are calculated using \textcolor{black}{ Eq.} \ref{eq:percentage_error}. The reference values for calculating the errors for the AE6 set are taken from Ref. \onlinecite{Lynch}. 
SCAN overestimates the \textcolor{black}{atomization energies slightly on average}, SCAN-SIC over-corrects and \textcolor{black}{hence underestimates them.} This is because, as we proceed from separated atoms to \textcolor{black}{the} molecules they form, the valence orbitals acquire more orbitals with which they overlap and, hence, become more noded. Thus, SIC to SCAN \textcolor{black}{makes the energy of the molecule higher than it would have been} if the orbital densities had been nodeless.\cite{Stretched} SCAN-sdSIC scales down \textcolor{black}{the SCAN-SIC results and yields an accuracy near that of SCAN. \textcolor{black}{The results for the total ground state energy of individual atoms and molecules using LSDA-LSIC, LSDA-LSIC+, SCAN, SCAN-sdSIC, and SCAN-SIC are presented in the supplemental information.}} 

\subsection{Chemical barrier heights}
\textcolor{black}{The} barrier height of a chemical reaction is the difference between the maximum energy of the transition state and the total energy of the reactants (forward barrier) or the products (reverse barrier). \textcolor{black}{The transition state involves} stretched bonds where SIE is expected to be \textcolor{black}{maximal}.\cite{Stretched} LSDA-LSIC and LSDA-LSIC+ both perform almost equally well for the representative set of barrier heights BH6.\cite{Lynch} The MAE of both the methods is better than that of SCAN-SIC and SCAN-sdSIC. 
Table \ref{tab:BH6} presents the MEs and MAEs of barrier heights for different approximations. The reference values are taken from Ref. \onlinecite{Grimme}.
 
\begin{table}[htbp]							
	\centering						
	\caption{Mean error (ME) and mean absolute error (MAE) of barrier heights for \textcolor{black}{the} BH6 set for various levels of approximation. The values are in kcal/mol.}					
	\begin{tabular}{ccc}						\hline\hline
 	{Approx.}	&	ME	&	MAE	\\	
 	\hline
 	LSDA	&	-18.1	&	18.1	\\	
 	LSDA-SIC	&	-5.1	&	5.1	\\	
 	LSDA-sdSIC	&	-4.1	&	4.1	\\	
 	LSDA-LSIC	&	0.6	&	1.4	\\	
 	LSDA-LSIC+	&	-0.2	&	2.1\\	
 	\multicolumn{1}{c}{}	&	&	\\			
 			
 	PBE	&	-9.6	&	9.6	\\	
 	PBE-SIC	&	0	&	4.6	\\	
 	PBE-sdSIC	&	-3.7	&	4.2	\\	
 	\multicolumn{1}{c}{}	&	&	\\			
 				
 	SCAN	&	-7.9	&	7.9	\\	
 	SCAN-SIC	&	-1	&	3	\\	
 	SCAN-sdSIC	&	-4.6	&	4.6	\\	
 	\hline\hline
	\end{tabular}%
	\label{tab:BH6}%
\end{table}%

\subsection{Ionization potential}
The ionization potential of a system refers to the energy required to remove the most loosely bound electron. \textcolor{black}{The} ionization potential can be calculated using either \textcolor{black}{the} vertical or \textcolor{black}{the} adiabatic method. Here, the adiabatic ionization potential of a system refers to the difference in energy of the neutral system, at its most stable \textcolor{black}{geometry,} and \textcolor{black}{of the cationic system,} at its stable geometry, whereas vertical ionization potential refers to the difference of the neutral and cationic system in the stable geometry of the neutral. \textcolor{black}{This means that, in the vertical} method, we assume that the geometry of the system remains unchanged even after the removal of an electron. \textcolor{black}{In the $\Delta SCF$} method, the ionization potential is obtained \textcolor{black}{as} the difference \textcolor{black}{between} the total ground state energy of the neutral and the cationic system. \textcolor{black}{The} vertical ionization potential can \textcolor{black}{also} be accessed by using Janak's theorem,\cite{Janak} \textcolor{black}{i.e., the derivative} of total energy with respect to orbital occupation is equal to the eigenvalue of that orbital $(\frac{\partial E}{\partial f_i}=\varepsilon_i)$, independent of the detailed form of the exchange-correlation functional. This implies \textcolor{black}{that the exact vertical ionization potential is simply the negative of the exact KS HOMO eigenvalue of} the neutral system.\cite{HOMO_ip,HOMO_ip2} Table \ref{tab:adiabatic_ip} shows the adiabatic ionization potential of \textcolor{black}{the} G2-1 set.\cite{G2,G2-1} For \textcolor{black}{the} adiabatic ionization potential with \textcolor{black}{the $\Delta SCF$} method, LSDA-LSIC+ improves over LSDA-LSIC and \textcolor{black}{is almost competitive} with SCAN and SCAN-sdSIC. \textcolor{black}{The} vertical ionization potential result of the same test set accessed using Janak's theorem is presented in Ref. \onlinecite{Santosh}. For a set of intermediate-sized organic molecules\cite{Santosh}, the error in the highest-occupied orbital energy is 0.83 eV for LSDA-LSIC+ and 1.13 eV for LSDA-LSIC.
\begin{table}[htbp]
  \centering
  \caption{Mean error (ME), and mean absolute error (MAE) of the adiabatic ionization potential of \textcolor{black}{the} G2-1 set using \textcolor{black}{the} $\Delta$SCF method for various levels of approximation. The reference values are taken from Ref. \onlinecite{G4}. The values are in kcal/mol.}
    \begin{tabular}{ccc}
    \hline\hline
    Method & ME    & MAE  \\
    \hline
    LSDA  & 2.7   & 6.2 \\
    LSDA-SIC & 7.0   & 10.3 \\
    LSDA-sdSIC & 3.6   & 7.8 \\
    LSDA-LSIC & 2.7   & 7.1 \\
    LSDA-LSIC+ & 1.4   & 6.4 \\
          &       &  \\
    PBE   & -0.8  & 4.8 \\
    PBE-SIC & -4.7  & 10.6 \\
    PBE-sdSIC & 0.8   & 6.8 \\
          &       &  \\
    SCAN  & -0.9  & 5.8 \\
    SCAN-SIC & -5.4  & 8.3 \\
    SCAN-sdSIC & -1.2  & 5.5 \\
    \hline\hline
    \end{tabular}%
  \label{tab:adiabatic_ip}%
\end{table}%
\subsection{Electron affinity}
\textcolor{black}{The} electron affinity of a neutral system is defined as the change in energy when an extra electron is added in its isolated gaseous phase. \textcolor{black}{Tables} \ref{tab:vertical_ea} and \ref{tab:adiabatic_ea}  show the MEs and MAEs of \textcolor{black}{the} electron \textcolor{black}{affinities of the G2-1} set\cite{G2,G2-1} using  vertical and adiabatic \textcolor{black}{methods}, respectively. \textcolor{black}{The adiabatic electron affinity assumes that the geometry of the system is relaxed after the addition of an electron.} In \textcolor{black}{the} vertical method, we assume that the geometry of the system remains unchanged even after adding an extra electron. \textcolor{black}{In the $\Delta SCF$ method, electron affinity is obtained as the difference between the total ground state energy of the neutral and anionic systems}. For an exact functional, \textcolor{black}{the negative of the eigenvalue} of the HOMO of an anionic system \textcolor{black}{yields} the vertical electron affinity.\cite{HOMO_ip,HOMO_ip2} We have accessed \textcolor{black}{the} vertical electron affinity of \textcolor{black}{the G2-1} set with this method. \textcolor{black}{Semi-local} DFAs fail to capture an extra electron in the anionic system. Hence, LSDA, PBE\textcolor{black}{,} and SCAN results presented here are single step non-SCF calculations with \textcolor{black}{the} SCAN-SIC density. LSIC and LSIC+ have equal MAEs for the adiabatic electron affinity and perform better than SCAN and SCAN-sdSIC. However, LSIC+ seems to be superior to LSIC for \textcolor{black}{the} vertical electron affinity of the G2-1 set. As single step non-SCF calculations do not produce eigenvalues, \textcolor{black}{the} quasi-SCF versions of LSIC and LSIC+ are used to calculate vertical electron affinity.

\begin{table}[htbp]
  \centering
  \caption{MEs and MAEs of \textcolor{black}{the} vertical electron affinity of \textcolor{black}{the} G2-1 set\textcolor{black}{,} calculated by using the \textcolor{black}{eigenvalue} of the HOMO. All values are in kcal/mol.}
  
    \begin{tabular}{ccc}
    \hline\hline
     Method     & ME    & MAE \\
     \hline
    LSDA   & -75.0 & 75.0 \\
    LSDA-SIC & 30.2  & 30.2 \\
    LSDA-LSIC & 6.5   & 8.9 \\
    LSDA-LSIC+ & -1.4  & 6.3 \\
    \hline\hline
    \end{tabular}%
  \label{tab:vertical_ea}%
\end{table}%

\begin{table}[htbp]
  \centering
  \caption{MEs and MAEs of the adiabatic electron affinity of \textcolor{black}{the} G2-1 set\textcolor{black}{,} calculated by \textcolor{black}{the} $\Delta SCF$ method. All values are in kcal/mol. The reference values are taken from Ref. \onlinecite{G2}. }
    \begin{tabular}{ccc}
    \hline\hline
     Method     & ME    & MAE \\
     \hline
    LSDA   & 5.7   & 5.9 \\
    LSDA-SIC & -1.4  & 5.6 \\
    LSDA-sdSIC & -0.3  & 3.4 \\
    LSDA-LSIC & 1.5   & 3.2 \\
    LSDA-LSIC+ & -0.2  & 3.2 \\
          &       &  \\
    PBE   & 1.3   & 2.0 \\
    PBE-SIC & -13.2 & 13.4 \\
    PBE-sdSIC & -4.5  & 4.6 \\
          &       &  \\
    SCAN  & -0.3  & 4.1 \\
     SCAN-SIC & -9.0  & 9.4 \\
     SCAN-sdSIC & -2.8  & 5.2 \\
   
    \hline\hline
    \end{tabular}%
  \label{tab:adiabatic_ea}%
\end{table}%

\subsection{Equilibrium bond length}
Table \ref{tab:Bondlength} shows the MEs and MAEs of \textcolor{black}{the} equilibrium bondlengths of \textcolor{black}{a} benchmark set of 11 diatomic molecules\cite{Vydrov}, evaluated by finding the \textcolor{black}{minimum} of the quadratic fit of the total energy over varying bond length. SCAN performs much better than any other functional. LSIC and LSIC+ values lie in between SCAN and SCAN-SIC, with LSIC having a slightly better result than the LSIC+ and closer to SCAN-sdSIC results.
\begin{table}[htbp]
  \centering
  \caption{MEs and MAEs of \textcolor{black}{the} equilibrium bond lengths of \textcolor{black}{a} benchmark set\textcolor{black}{\cite{Vydrov}} of  11 diatomic molecules (in angstrom). Results other than those for LSIC+ are the same as in Table VIII of Ref. \onlinecite{Puskar}, where the large improvement due to interior-scaling of the SIC was first found.}
    \begin{tabular} {ccc}
    \hline\hline
    Method & ME    & MAE \\
    \hline
    LSDA   & 0.0076 & 0.0110 \\
    LSDA-SIC & -0.0317 & 0.0392 \\
    LSDA-sdSIC & -0.0085 & 0.0189 \\
    LSDA-LSIC & -0.0015 & 0.0129 \\
    LSDA-LSIC+ & -0.0024 & 0.0147 \\
          &       &  \\
          &       &  \\
    PBE   & 0.0123 & 0.0123 \\
    PBE-SIC & -0.0134 & 0.0257 \\
    PBE-sdSIC & -0.0019 & 0.0132 \\
          &       &  \\
          &       &  \\
    SCAN  & 0.0039 & 0.0057 \\
    SCAN-SIC & -0.0190 & 0.0197 \\
    SCAN-sdSIC & -0.0110 & 0.0110 \\
    \hline\hline
    \end{tabular}%
  \label{tab:Bondlength}%
\end{table}%
\subsection{Dissociation energy of noncovalently bonded complexes}
\textcolor{black}{The S22 dataset\cite{S22} consists of the dissociation energies of complexes built up from pairs of small to large, mostly organic molecules, at reference geometries. These complexes are divided into three categories: hydrogen bonded complexes, complexes with predominant dispersion interactions, and mixed complexes (with both kinds of bonds). We tested the performance of different approximations on this dataset. Table \ref{tab:S22} presents the MEs and MAEs in kcal/mol for these functionals.  All of our calculations in Table \ref{tab:S22}} \textcolor{black}{are spin unpolarized} \textcolor{black}{and self-consistent,  with the exception of LSDA-LSIC and LSDA-LSIC+ (evaluated on LSDA-SIC total and localized orbital densities) and SCAN-sdSIC (evaluated on SCAN-SIC total and localized orbital densities). While the other functionals perform reasonably well, both interior-scaled functionals, LSDA-LSIC and LSDA-LSIC+, \textcolor{black}{consistently underestimate the binding of the complexes, in several cases predicting them to be unbound.}
Detailed results for the individual molecules are presented in the supplemental information.}
\textcolor{black}{
Until we can implement LSDA-LSIC and LSDA-LSIC+ self-consistently, we cannot say whether their failures for weak bonds might be corrected by full self-consistency. All we can say is that the LSDA-SIC and SCAN-SIC densities and FLO densities on which they are evaluated are not by themselves wrong, since the LSDA-SIC and SCAN-SIC S22 binding energies are reasonable.}

\textcolor{black}{It might be surprising that the original PZ SIC (labelled SIC in Table \ref{tab:S22}) is not worse than it is for the S22 set, since the self-interaction corrections from the valence orbitals are large in comparison with the weak binding energies of the S22 complexes.  The explanation for this lies in the chemical interpretation of the FLOs as covalent double bonds, covalent single bonds, lone pairs, and core orbitals. These orbitals are often changed only modestly when weak bonds are broken. Thus most of the self-interaction correction from the valence electrons cancels out of the binding energies of the complexes. The cancellation is even more perfect when the weak bonds are not all broken but are simply re-arranged, as in the structural energy differences among the water hexamers\cite{FLOSIC_app6}.  An exception\cite{Kamal} to the partial cancellation upon weak-bond breaking is the hydrogen bond in $H_3O^+(H_2O)$, which is strong enough to induce a transition of electrons between covalent-bond orbitals and lone-pair orbitals.}

\textcolor{black}{The cancellation in PZ SIC discussed in the previous paragraph can be lost in LSDA-LSIC (evaluated on LSDA-SIC densities and orbitals). The FLO densities are what they were in LSDA-SIC, but their contributions to the self-interaction correction are now scaled down, and the scaling \textcolor{black}{can be} stronger in the overlap region than in the corresponding regions of the same monomer or isolated molecule.  This arises because of a reduction in the scaling parameters $z_\sigma$ and $f(z_\sigma)$ in the region where the charge from the two components in a given complex overlaps.}  

\begin{figure}[h!]
    \centering
    \includegraphics[scale=0.5]{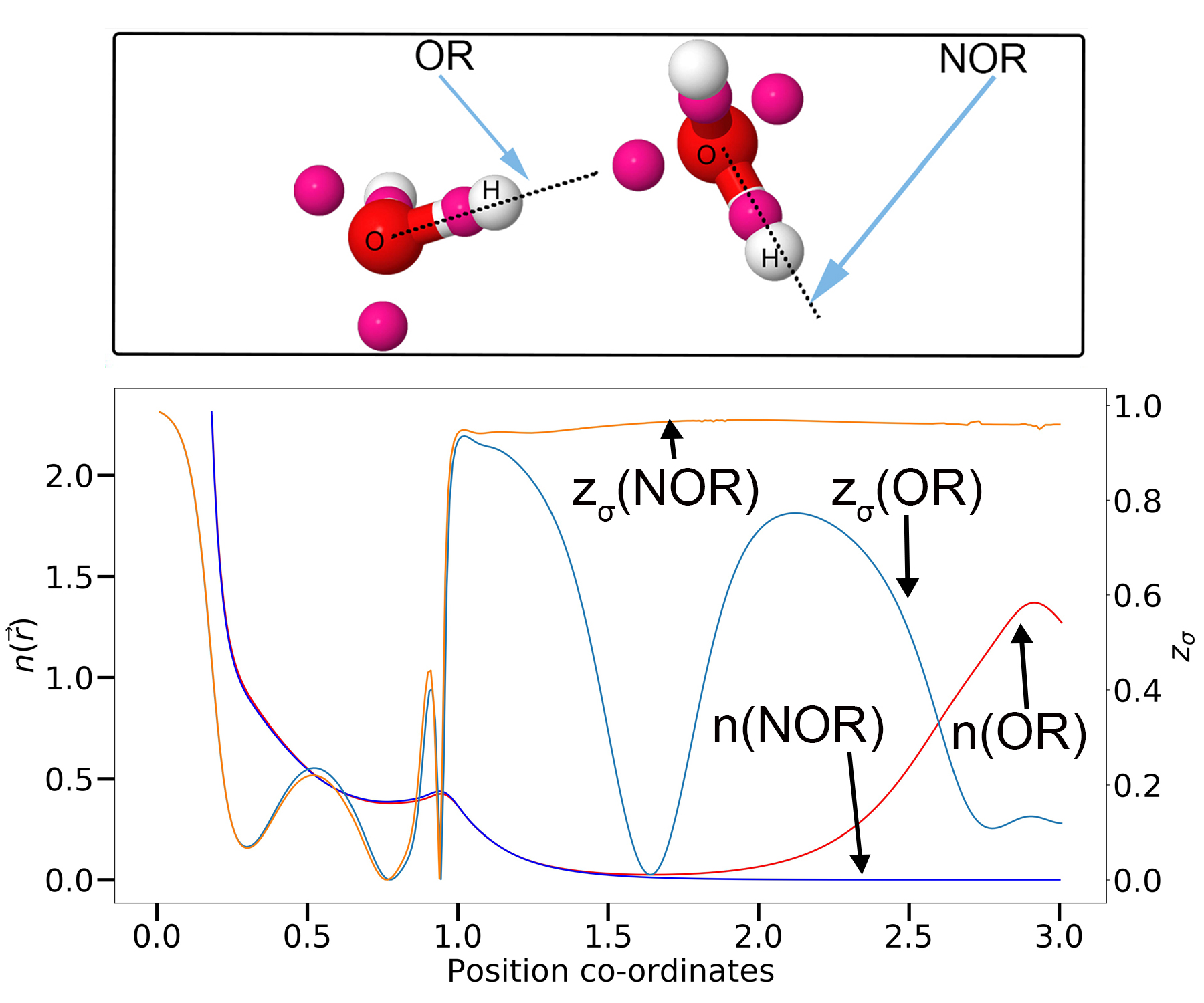}
    \caption{\textcolor{black}{Comparison of the total electron density $n$ and the scaling factor $z_\sigma$ in the overlapping region (OR) and a corresponding non-overlapping region (NOR) of the water dimer.  The position of the O atom is at the origin for each curve. In the inset is a diagram that shows the positions of the O (red) and H (white) atoms, as well as the FOD positions (pink). The scaling factor and the total electron density are plotted along the dotted lines shown on the inset. The position co-ordinates are in \AA \ and the total electron densities are in atomic units.}}
    \label{fig:water_dimer}
\end{figure}

\textcolor{black}{In $(H_2O)_2$, for example, the sum of the orbital self-interaction corrections to the total energy differs by only 0.2 mHa from the sum of the corresponding SIC orbital energies in the two isolated molecules.  In the LSIC calculation, however, this difference is 10.0 mHa (about 6 kcal/mol).  This difference can be traced to contributions from FLOs corresponding to the O lone-pair orbital and the O-H bond orbital that are located in the overlap region between the two water molecules. \textcolor{black}{Fig. \ref{fig:water_dimer} shows the comparison of the total electron density $n$ and the scaling factor $z_\sigma$ of the water dimer along the bond axis of O-H in the overlapping region and the corresponding non-overlapping region. The plot omits the very large density near the O atom nucleus. It can be seen that the value of $z_\sigma$ as well as $n$ remains similar up to the position of the H atom at about 0.96 \AA. The density remains almost equal until about 1.5 \AA \ but $z_\sigma$ differs significantly. After about 1.5 \AA, the density in the overlapping region increases because of the presence of the lone pair electron of another water molecule.} The scaling factor $z_\sigma \ (\ f(z_\sigma)$ for LSIC+) is seen to be smaller in the overlap region than at corresponding locations outside the overlap region or in the isolated molecules. Because of this, the LSIC energies for these orbitals are relatively smaller in magnitude than the LSIC energies for the corresponding orbitals in the isolated molecules, resulting in the energy difference noted above.  Because the LSDA-SIC and LSDA-LSIC orbital energies are negative, and the energy densities that get scaled down are primarily negative, the implication is that the complex is relatively less stable in LSDA-LSIC than in LSDA-SIC.  Similar effects are found for other complexes we have analyzed.} 

\textcolor{black}{Why is the scaling factor smaller in the overlap region?  $z_\sigma$ is close to one in one-electron regions of the charge density.  Far from the center of an isolated molecule, the density tends to be dominated by contributions from a single orbital and $z_\sigma$ is correspondingly close to one.  In the overlapping charge region of the molecular complex, however, the density decreases to a minimum moving away from one molecule and then increases again on nearing the second molecule.  Since $z_\sigma$ depends on the gradient of the density (see Eq. \ref{eq:tau_sigma}), the scaling factor must be close to zero near this minimum.}

\textcolor{black}{Performing self-consistent LSDA-LSIC and LSDA-LSIC+ calculations could improve the results for the S22 interaction energies. Shifting the electron density in the overlap region of the complexes slightly could have a significant impact on the scaling parameters $z_\sigma$ and $f(z_\sigma)$ without changing other quantities that affect the total energy. Alternatively, employing LSDA-LSIC and LSDA-LSIC+ -type approaches with scaling functions based on improved iso-orbital indicators $\alpha\cite{TPSS}$ or $\beta\cite{Furness}$ could also improve the results for the interaction energies.  Unlike $z_\sigma$, these indicators can distinguish densities associated with weak (i.e., van der Waals or hydrogen) bonds from slowly-varying densities.}
\begin{figure}[h!]
    \centering
    \includegraphics[scale=0.23]{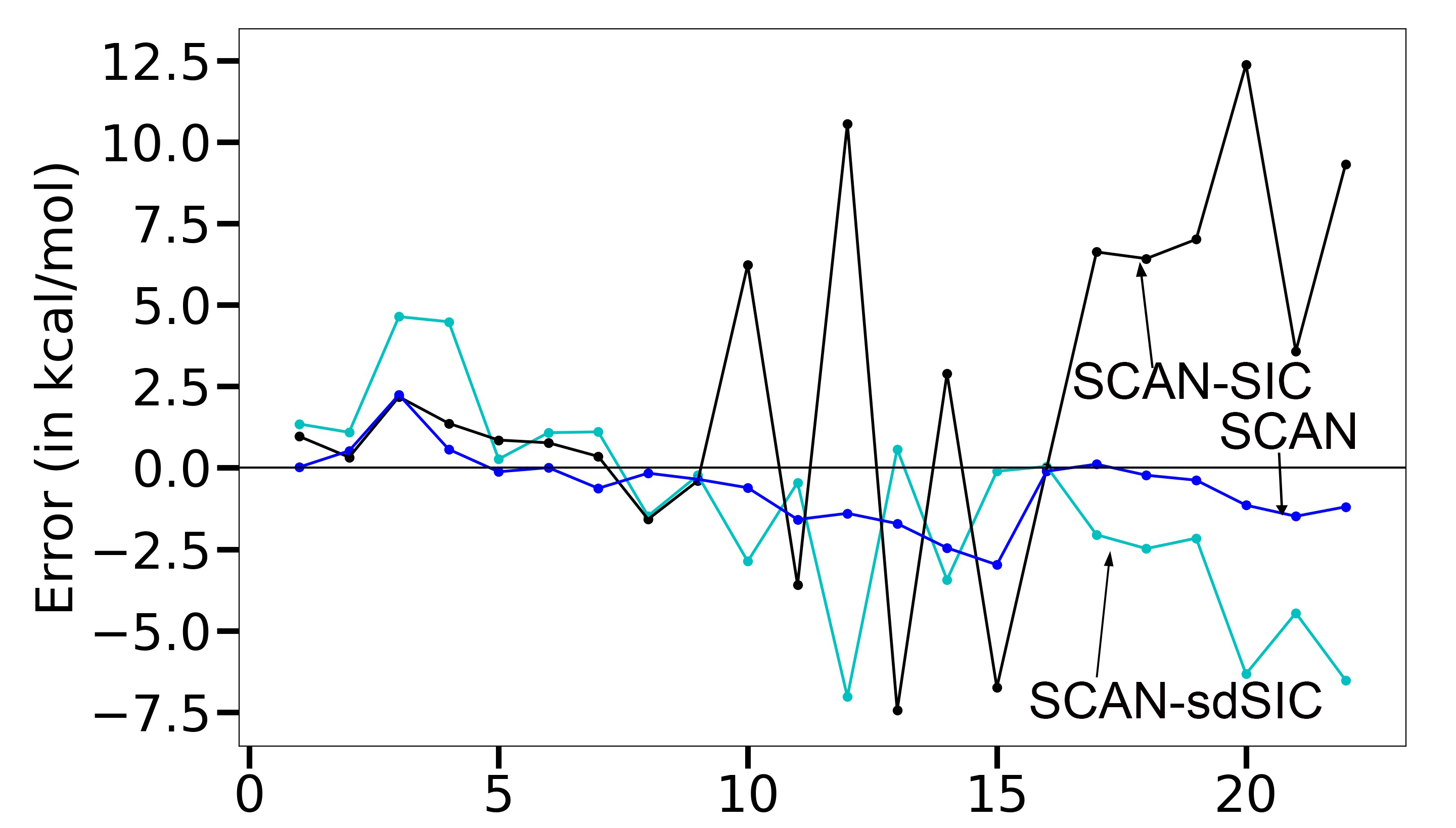}
    \caption{ Individual errors(in kcal/mol) for the dissociation energy of complexes of \textcolor{black}{the} S22 dataset}
    \label{fig:S22_individual}
\end{figure}

\begin{figure}[h!]
         \includegraphics[scale=2.75]{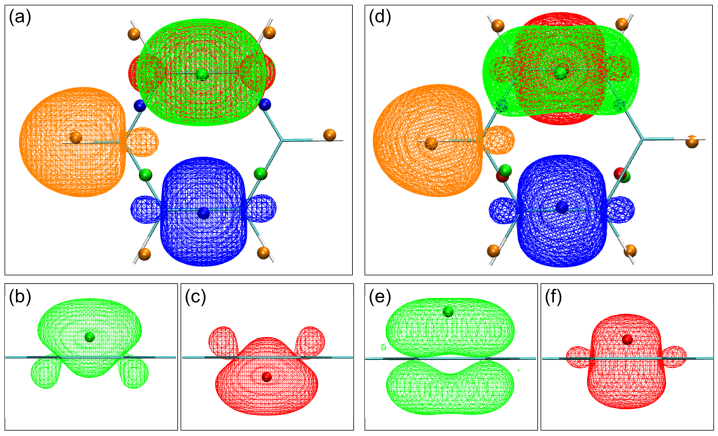}
       
    \caption{FLO densities of the benzene molecule obtained using two FOD arrangements. The \textit{sym} arrangement yields the FLOs in panel (a) (top view) and (b) and (c) (side views). The \textit{asym} arrangement gives the FLOs in panel (d), (e) and (f). FLOs corresponding to C-H bonds (orange), C-C single bonds (blue), and C-C double bonds (red and green) are shown.  Isosurface values of 0.001 $e/$\AA$^3$ were used to make all FLO images. FODs corresponding to each FLO type are shown by small spheres of the corresponding color. In the side views, only the double bond FLOs are shown along with  the corresponding FOD positions.}
    \label{fig:benzene_dimer}
\end{figure}

 \begin{table}[h!]
  \centering
  \caption{The dissociation energy of weakly bound complexes in the S22 dataset. All values are in kcal/mol. The calculations are 
spin-restricted. The reference values are taken from Ref. \onlinecite{s22_ref}. A negative ME indicates that an approximation 
underestimates the binding, on average.}
    \begin{tabular}{ccc}
    \hline\hline
          & ME    & MAE \\
          \hline
    LSDA   & 2.16 & 2.18 \\
    LSDA-SIC & 1.41 & 2.98 \\
    LSDA-LSIC & -8.95 & 9.77 \\
    LSDA-LSIC+ & -7.00 & 7.42 \\
          &       &  \\
    PBE   & -2.56 & 2.58 \\
    PBE-SIC & -2.64 & 2.97 \\
          &       &  \\
    SCAN  & -0.60 & 0.91 \\
    SCAN-SIC & 2.37 & 4.16 \\
    SCAN-sdSIC  & -1.14 & 2.46 \\
    \hline
    \hline
    \end{tabular}%
  \label{tab:S22}%
\end{table}%
\textcolor{black}{Another conclusion from Table \ref{tab:S22} and Ref. \onlinecite{Puskar} is that SCAN-sdSIC, evaluated on SCAN-SIC densities and FLOs, is almost uniformly more accurate than SCAN-LSIC or SCAN-LSIC+ evaluated on LSDA-SIC densities and FLOs for equilibrium properties of molecules, and especially so for the weakly-bound complexes. Figure \ref{fig:S22_individual} shows the individual errors in the dissociation energy of the complexes in the S22 database by SCAN, SCAN-SIC, and SCAN-sdSIC.  SCAN-sdSIC overestimates and makes slightly more error than both SCAN and SCAN-SIC for hydrogen-bonded complexes (complexes 1-7).  SCAN-sdSIC restores the capacity of SCAN to capture the short and intermediate range van der Waals interactions involved in the complexes with predominant dispersion contribution (complexes 8-15) and the mixed complexes (complexes 16-22) which are somewhat degraded from SCAN to SCAN-SIC. Hence, SCAN-sdSIC might be a better option to scale down the PZ-SIC and predict the properties of weakly bonded systems.}

\textcolor{black}{\subsection{Do the FLOs have a consistent chemical interpretation?}} 
\textcolor{black}{The FLOs are often interpreted chemically as \textcolor{black}{covalent double bonds, covalent single bonds,} lone pairs, etc.}  \textcolor{black}{But this interpretation is only useful if it can be made consistently. The benzene molecule is a challenging example in this context.  There are two distinct solutions for the spin unpolarized benzene monomer in LSDA-SIC.  In one (\textit{sym}), the FODs are as shown in Fig. 8:  for the C-C bonds around the ring, a single FOD at the center of one bond alternates with a pair of FODs that are symmetrically placed above and below the bond center of the next, nominally representing single (C-C) and double (C=C) bonds, respectively.  The single bond FOD gives rise to the FLO density depicted in blue in Fig. 8, while the double bond FODs give rise to a pair of symmetric FLOs, one above the plane of the molecule and the below, its mirror image, below it (Fig. 8 (b) and (c)).  
In the other solution (\textit{asym}), the double bond FODs both lie on the same side of the molecular plane.   In this case, the single-bond FLO is essentially unchanged from the \textit{sym} solution, but the double bond FLOs change shape dramatically to the red and green FLO densities shown in Fig. 8 (e) and (f). } \textcolor{black}{(Note that it is the FODs, and not the FLO densities, that can be strongly asymmetric around the plane of the molecule.)}  \textcolor{black}{  The red FLO is very similar to the single bond FLO and can be thought of as a bonding combination of $sp^2$ hybrid orbitals.  The green FLO can be thought of as a bonding combination of $p_z$ orbitals and has a node in the molecular plane.  In the \textit{sym} solution, the symmetric double bond FLOs each have an SIC energy of -0.016 Ha.  In the \textit{asym} solution, the red and green FLOs have SIC energies of -0.048 and +0.023 Ha, respectively.  Despite this, the total energy of the \textit{asym} solution is only 1.4 mHa (0.9 kcal/mol) lower in energy than the \textit{sym} solution in LSDA-SIC.  This remarkable near degeneracy is broken in LSDA-LSIC, where the ordering switches and the \textit{sym} solution is lower by 23.2 mHa (14.6 kcal/mol).  In the benzene-containing complexes, we only found \textit{asym}-type LSDA-SIC solutions, despite using \textit{sym}-type starting points for the FODs.  The \textit{asym} and \textit{sym} solutions are also nearly degenerate in PBE-SIC.}

\textcolor{black}{There is also a third nearly degenerate solution for the benzene monomer if spin polarized densities are considered.  This solution (\textit{Linnett}) has been discussed elsewhere\cite{Schwalbe_FOD} and can be linked to the Linnett double quartet bonding theory,\cite{Linnett_1,Linn2} a spin-polarized generalization of Lewis theory.  In this case, \textit{sym}-type FOD arrangements in the up and down spin channels are shifted by one bond around the ring, so that each bond has a single FOD of one spin at the bond center and two FODs of the opposite spin placed symmetrically above and below the center.  The FLOs for the \textit{Linnett } solution are very similar to the \textit{sym} FLOs shown in Fig. 8.  The total energy of the \textit{Linnett} solution in LSDA-SIC is slightly lower (1.3 mHa = 0.8 kcal/mol) than that of the \textit{asym} solution.  For the parallel-displaced benzene dimer (S22 complex 11), we find that the \textit{Linnett} solution for the dimer is also slightly lower than the \textit{asym} dimer, so that the calculated dissociation energies are nearly identical, 1.6 kcal/mol and 1.1 kcal/mol for \textit{Linnett} and \textit{asym} respectively.  The data in Table X reflect spin unpolarized calculations only.  We do not expect that significant differences would arise if spin-polarized \textit{Linnett} arrangements were used for all complexes containing benzene.
}


\textcolor{black}{The way to find the localized orbitals suggested by Perdew and Zunger 1981, and universally applied since then, is to choose the unitary transformation of the canonical or generalized Kohn-Sham orbitals that minimizes the self-interaction-corrected total energy. But that approach is inconsistent with the chemical interpretation of the FLOs. To preserve \textcolor{black}{a consistent} chemical interpretation, we need a new way to select the optimum unitary transformation, for example, by minimizing a measure of the inhomogeneity of the FLO densities.  Such a measure should penalize all strongly noded or lobed orbital densities such as the green ones in Fig. 8 (d) and (e).}

\textcolor{black}{The nuclear geometry used here for benzene is unchanged under 60-degree rotations around a central axis. There must be an exact ground-state electron density that preserves this symmetry.\cite{PRNK} The single-determinant auxiliary wave functions that we have calculated break this symmetry. In either the spin unpolarized \textit{sym}/\textit{asym} or the \textit{Linnett} scheme, there are two degenerate Slater determinants with different densities (\textit{sym}/\textit{asym}) or spin densities (\textit{Linnett}), the density or spin density of one determinant transforming into that of the other under a 60-degree rotation.  A possible physical interpretation\cite{PRNK} is that there are low-frequency fluctuations of the density and spin-density that anti-correlate on alternate bonds around the benzene ring.}
\\
\section{Summary and conclusions}
We introduced LSIC+ that \textcolor{black}{takes the} iso-orbital indicator $z_\sigma$ as its argument for interior scaling of Perdew-Zunger self-interaction correction. LSIC+ is applied with LSDA only and not with GGAs and meta-GGAs to avoid gauge-inconsistency. LSIC+ is an improved version of LSIC\cite{Zope} that satisfies one more constraint \textcolor{black}{from} the large-$Z$ asymptotic expansion of exchange-correlation energy\textcolor{black}{,} in addition to restoring the correct uniform density limit of \textcolor{black}{the} exchange energy violated by PZ-SIC. The relevance of the uniform-density and large-Z limits to valence-electron properties was explained in Ref.~\citenum{Kaplan}.  LSIC+ retains the full PZ-SIC in the region of one-electron density, \textcolor{black}{giving} no correction for uniform electron density, and a scaled correction in the region in between. Hence, LSDA-LSIC+ works well for both stretched bond cases and equilibrium properties that do not involve weak bonds. We present here the results of LSDA-LSIC+ for \textcolor{black}{ground state energies of atoms from \textcolor{black}{$Z=$1-18,} atomization energies of the AE6 set, chemical barrier heights of the BH6 set, ionization potentials of the G2-1 set, electron affinities of the G2-1 set, equilibrium bond lengths, and interaction energies} of \textcolor{black}{the} S22 set of \textcolor{black}{weakly-bonded complexes,} and compare with LSDA-LSIC and other \textcolor{black}{SIC-corrected} and uncorrected DFAs. LSDA-LSIC+ performs better than LSDA-LSIC for most of the properties and \textcolor{black}{comparably} for a few. LSDA-LSIC+ is even better than SCAN or SCAN-SIC for a few properties such as total ground state energy of atoms and electron affinity.
LSIC+ satisfies one more constraint than LSIC and produces better results than LSIC, \textcolor{black}{which} shows the significance of constraint satisfaction in the construction of functionals. The better performance of SCAN-sdSIC can also be attributed to the recovery of \textcolor{black}{higher-order} coefficients \textcolor{black}{in the asymptotic expansion of exchange-correlation energy for atoms of large $Z$.} However, since LSIC and LSIC+ \textcolor{black}{employ} functions of $z_\sigma$, they \textcolor{black}{cannot} recognize\cite{weak_bonds} weaker bonds and \textcolor{black}{(at least when evaluated on LSDA-SIC total and localized orbital densities)}produce \textcolor{black}{poor results for the van der Waals or hydrogen-bond binding energies.}  

\textcolor{black}{Surprisingly, the Fermi-L\"{o}wdin orbitals that describe multiple covalent bonds are not unique.  Two different versions give nearly degenerate LSDA-SIC energies in benzene-like systems, though not in LSDA-LSIC.  A more physical way to determine the Fermi orbital descriptors is needed to give a single, consistent picture of these FLOs.}

Unlike LSIC and LSIC+, SCAN-sdSIC is applied to \textcolor{black}{the} more accurate SCAN functional, which is already much better than LSDA and hence \textcolor{black}{improves its predictive power.} It provides good results for several ground state properties discussed here, including the dissociation energy of weakly bonded systems.  
The SCAN-sdSIC, if scaled down by functions of improved iso-orbital indicators $\alpha$ or $\beta$, may \textcolor{black}{further} improve the results.\cite{SOSIC} However, since SCAN-sdSIC provides an incorrect \textcolor{black}{asymptote} $\frac{-X_{HO}}{r}$ \textcolor{black}{for the} exchange-correlation potential, it \textcolor{black}{does not} guarantee a good description of charge transfer. \textcolor{black}{The optimal SIC that remains to be developed might be PZ SIC evaluated on complex\cite{complex_orbital_importance} Fermi-Löwdin orbitals\cite{Pederson} (with nodeless\cite{Stretched,complex_orbital_importance} orbital densities) and with Fermi orbital descriptors chosen to minimize a measure of the inhomogeneity of the orbital densities.}  


\begin{acknowledgments}
This work was supported by the U.S. Department of Energy, Office of Science, Office of Basic Energy Sciences, as part of the Computational Chemical Sciences Program under Award No. DE-SC0018331. \textcolor{black}{The work of P.B. and KW. was supported by the U.S. National Science Foundation under Grant No. DMR-1939528.} This research includes calculations carried out on HPC resources supported in part by the National Science Foundation through major research instrumentation grant number 1625061 and by the US Army Research Laboratory under contract number W911NF-16-2-0189.  A few calculations \textcolor{black}{were} carried out on the EFRC cluster supported by the Department of Energy (DOE), Office of Science (OS), Basic Energy Sciences (BES) through Grant No. DE-SC0012575 to the Energy Frontier Research Center: Center for Complex Materials from First Principles. The plots are generated by MATPLOTLIB.\cite{matplotlib}

\vspace{0.5cm}
\end{acknowledgments}

\section{Supplemental material}
The detailed results of ground state energy of atoms \textcolor{black}{($Z=$1-18)}, \textcolor{black}{the} energy of individual atoms, and molecules of \textcolor{black}{the} AE6 set and \textcolor{black}{the} S22 dataset can be accessed through supplementary material. Other data in detail that support the findings of the study are available from the corresponding author upon reasonable request. 
\begin{appendices}
\appendix
\end{appendices}

\bibliographystyle{aipnum4-1}
\end{document}